\documentclass[letterpaper]{article} % DO NOT CHANGE THIS
\usepackage{aaai2026}  % DO NOT CHANGE THIS

\usepackage{times}  % DO NOT CHANGE THIS
\usepackage{helvet}  % DO NOT CHANGE THIS
\usepackage{courier}  % DO NOT CHANGE THIS
\usepackage[hyphens]{url}  % DO NOT CHANGE THIS
\usepackage{graphicx} % DO NOT CHANGE THIS
\usepackage{natbib}  % DO NOT CHANGE THIS AND DO NOT ADD ANY OPTIONS TO IT
\usepackage{caption} % DO NOT CHANGE THIS AND DO NOT ADD ANY OPTIONS TO IT
\usepackage{algorithm}
\usepackage{algorithmic}

\usepackage{multirow}
\usepackage{xcolor}
\usepackage{booktabs}
\usepackage{subcaption}
\usepackage{float} 
\usepackage{amsfonts}
\usepackage{mathrsfs}
\usepackage{amsmath}

\usepackage{adjustbox}

\newcommand{\eg}{\textit{e}.\textit{g}.}
\newcommand{\ie}{\textit{i}.\textit{e}.}

\usepackage{newfloat}
\usepackage{listings}
\DeclareCaptionStyle{ruled}{labelfont=normalfont,labelsep=colon,strut=off} % DO NOT CHANGE THIS
\floatstyle{ruled}
\newfloat{listing}{tb}{lst}{}
\floatname{listing}{Listing}
\title{LLM-I2I: Boost Your Small Item2Item Recommendation Model with Large Language Model}
\author{
    Yinfu Feng\textsuperscript{\rm 1}\equalcontrib,
    Yanjing Wu\textsuperscript{\rm 1}\equalcontrib,
    Rong Xiao\textsuperscript{\rm 1},
    Xiaoyi Zeng\textsuperscript{\rm 1}
}
\affiliations{
    \textsuperscript{\rm 1} Alibaba Group\\
    Hangzhou, Zhejiang, China\\
    fyf200502@gmail.com, qinge.wyj@alibaba-inc.com, \{xiaorong.xr, yuanhan\}@taobao.com
}

\usepackage{bibentry}
\begin{document}
% \begin{sloppypar}      % 设置自动换行，文章两端对齐

\maketitle

\begin{abstract}
Item-to-Item (I2I) recommendation models have become a cornerstone of many real-world recommendation systems owing to their great scalability, real-time recommendation capabilities, and relatively high recommendation quality. To improve their performance, there are two main research directions: 1) model-centric methods, update these small and shallow models into deeper models, and 2) data-centric methods, refine or synthesize much more high-quality data. The former approaches would introduce risks related to model changes, computation resource utilization, and online response times. In contrast, the latter approaches, which do not alter the models, avoid increases in online deployment and service resource requirements, making it more cost-effective in practice applications. However, data sparsity and noise problems often affect their performance. In light of this, we pay attention to data-centric methods in this paper and propose a \textbf{L}arge \textbf{L}anguage \textbf{M}odels enhanced \textbf{I}tem-to-\textbf{I}tem method (\textbf{LLM-I2I}) to overcome data sparsity and noise problems. Specifically, we first learn an LLM-based data generator with user historical behavior data to synthesize some user-item interaction data, especially for long-tail items, to alleviate the data sparsity problem in the originally collected data. Then, we build an LLM-based data discriminator to filter out weak or noisy user-item interactions, refining the collected and synthetic data. After that, the refined real and synthetic data are fused for training I2I models. To evaluate the proposed LLM-I2I, we apply it to various I2I models and compare their performance on industrial and academic datasets (\ie  AEDS and ARD). Experimental results indicate that LLM-I2I can significantly improve recommendation results, especially for long-tail items. In addition, deploying LLM-I2I to a large-scale cross-border e-commerce platform leads to 6.02\% and 1.22\% improvements over the existing I2I-based model in terms of recall number (RN) and gross merchandise value (GMV), respectively.    
\end{abstract}


\section{Introduction}
Item-to-item (I2I) recommendation algorithms have become fundamental components in modern recommendation systems across various domains, including e-commerce, news, and video platforms\cite{linden2003amazon,garcin2012personalized,chen2017attentive,bell2007improved,islam2021debiasing}. These systems typically operate in two stages: offline model training and online recommendation generation. During the offline phase, the system analyzes user-item interactions (e.g., clicks, purchases) to establish item\footnote{In e-commerce, we refer to a product as a unique document, often using the terms "document" or "item" when there is no confusion.} similarity relationships. In the online phase, these relationships enable real-time personalized recommendations by matching user-interacted items with similar candidates. Unlike the popular deep-mode-based recommendation algorithms, these lightweight and smaller I2I recommendation models require minimal offline and online computing resources. These characteristics make them particularly valuable for resource-constrained scenarios and large-scale applications. Despite the emergence of numerous advanced recommendation algorithms, I2I models maintain their practical significance in industrial systems, warranting continued research and optimization efforts.

\begin{figure}[htbp]
	\centering
	\begin{subfigure}{\linewidth}
		\centering
		\includegraphics[width=0.9\linewidth]{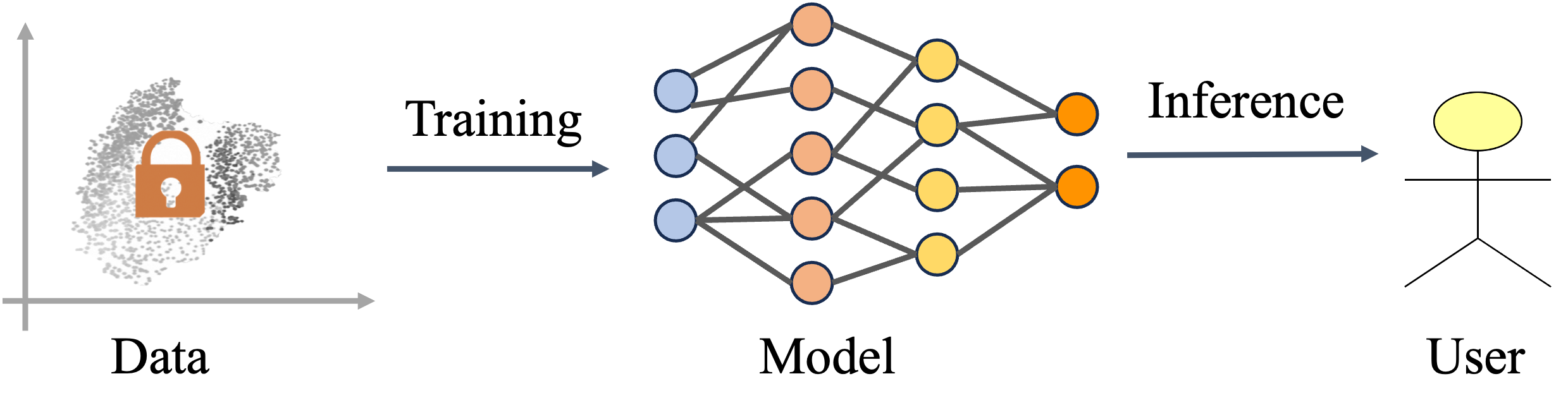}
		\subcaption{Model-centric optimization: Refining recommendation model architecture and learning algorithms.}
		%\label{model_centric}
	\end{subfigure}
	\begin{subfigure}{\linewidth}
		\centering
		\includegraphics[width=0.9\linewidth]{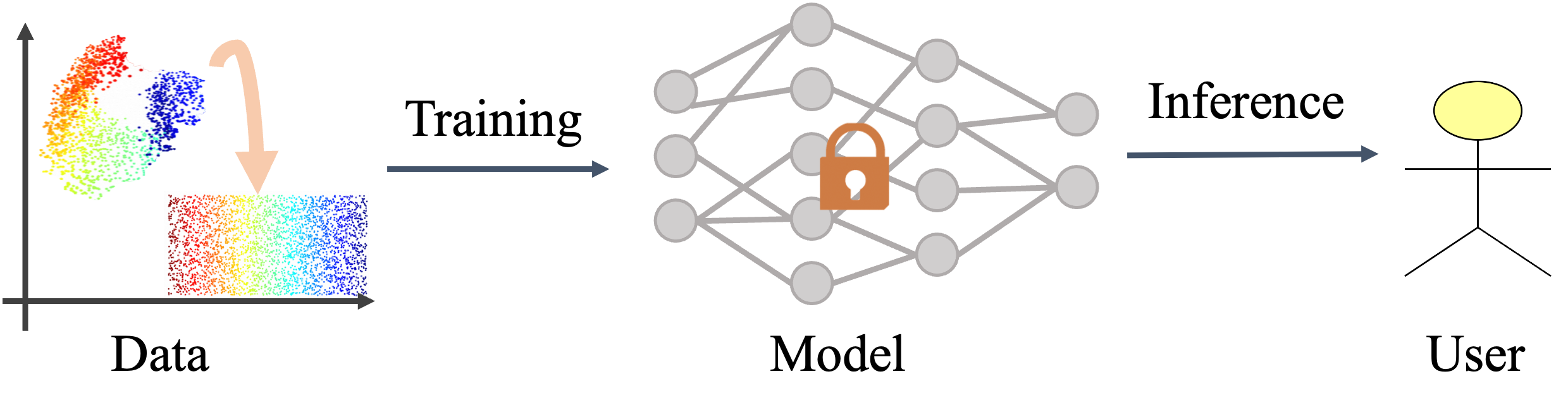}
		\subcaption{Data-centric optimization: Enhance data quality, feature engineering and user behavior modeling.}
		%\label{data_centric}
	\end{subfigure}
        % \caption{Model-centric methods（a） vs data-centric methods (b)}
        \caption{A comparative illustration of model-centric vs. data-centric approaches in recommendation system optimization.}
        \label{fig_data_centric_and_model_centric}
\end{figure}

I2I model enhancement primarily follows two research directions (Figure \ref{fig_data_centric_and_model_centric}): model-centric and data-centric methods\cite{Yin_2024}. Model-centric methods focus on developing deeper architectures to improve recommendation capabilities \cite{xue2019deep}, often at the cost of increased computational requirements. In contrast, data-centric methods employ various augmentation techniques to enhance training data quality and diversity \cite{sun2024large}. These methods offer distinct advantages: minimally impact online computational resources, require no modifications to existing models, maintain real-time response capabilities, and provide cost-effective implementation. Given these benefits, we focus on data-centric methods, particularly preserving existing system infrastructure and real-time performance.

Data-centric approaches can be classified into heuristic-based and model-based methods \cite{dang2024data}. Heuristic-based techniques employ predefined rules or random operations for data augmentation, generating new samples by modifying existing data patterns without requiring model training \cite{tan2016improved,zhou2023equivariant}. In contrast, model-based methods utilize trainable augmentation modules that learn to generate data according to specific objectives or constraints. This ability to incorporate targeted learning objectives makes model-based approaches generally more effective than their competitors.

Model-based data augmentation approaches can be divided into three groups: \textit{Sequence Extension and Refining}, \textit{Sequence Generation}, and \textit{LLM-Based Augmentation}\cite{dang2024data}. \textit{Sequence Extension and Refining} focuses on enhancing individual user behavior sequences through prediction or decision modules. \textit{Sequence Generation} aims to understand and replicate the overall data distribution by generating new data points within learned latent spaces or capturing underlying data patterns. \textit{LLM-Based Augmentation} combines human-provided instructions with LLM's world knowledge to generate enhanced data. This method has demonstrated effectiveness in various recommendation scenarios. For instance, LLM-CF \cite{sun2024large} advances this approach through synthesized In-Context COT datasets. Despite significant advancements, LLM-based augmentation still faces the following issues:

\textbf{Sparsity Problem}. In the realm of e-commerce recommender systems, data sparsity poses a significant challenge, particularly for long-tail items. These items are often difficult to recommend due to historical user-item interaction records scarcity. To address this issue, \citeauthor{wang2024large} proposed a prompt construction method, which has shown some efficacy in alleviating the issue. However, this method is primarily limited to small-scale cold-start items. It is unsuited for real-world recommendation systems that handle tens of millions of long-tail items. Moreover, while content-based recommendation methods can aid in suggesting long-tail items, their effectiveness is constrained by the limitations of traditional shallow or deep algorithmic models.

\textbf{Noise Problem}. In many real-world applications, the datasets used for training algorithmic models are often complex and require careful preprocessing. Specifically, in e-commerce recommendation systems, user-item interaction data can be noisy and messy. For example, users may accidentally click on items, leading to spurious interaction records. Training an I2I recommendation model solely on raw user-click data without proper filtering can result in inaccurate recommendations. While LLM-based data augmentation methods can increase the volume of training data, they may also introduce inaccuracies. According to \citeauthor{wang2024llm4dsr}, relying solely on synthetic data for training without appropriate filtering can diminish the effectiveness of recommendations. Therefore, it is crucial to carefully design data augmentation and filtering strategies to balance the trade-off between data quantity and quality, especially when dealing with long-tail items and cold-start problems in recommendation systems.

To address these problems, we propose a method named \textbf{L}arge \textbf{L}anguage \textbf{M}odels enhanced \textbf{I}tem-to-\textbf{I}tem (LLM-I2I). Our main contribution can be summarized as follows:
\begin{itemize}
    \item A new LLM-enhanced I2I model (LLM-I2I) has been proposed and has made significant progress in enhancing the widely used traditional I2I models.
     
    \item Unlike existing works that often overlook the combination of data generation and selection, LLM-I2I integrates an LLM-based data generator with a data discriminator to address data sparsity and noise issues effectively.

    \item Extensive experiments conducted on industry and academic datasets show that LLM-I2I significantly enhances the performance of various existing I2I models. Additionally, we successfully deployed the proposed LLM-I2I on a large-scale online e-commerce platform (i.e., AliExpress.com), obtaining 6.02\% and 1.22\% improvements in recall number (RN) and gross merchandise value (GMV) separately, highlighting its practical and commercial value. 
\end{itemize}

\begin{figure*}
    \centering
    \includegraphics[width= \linewidth]{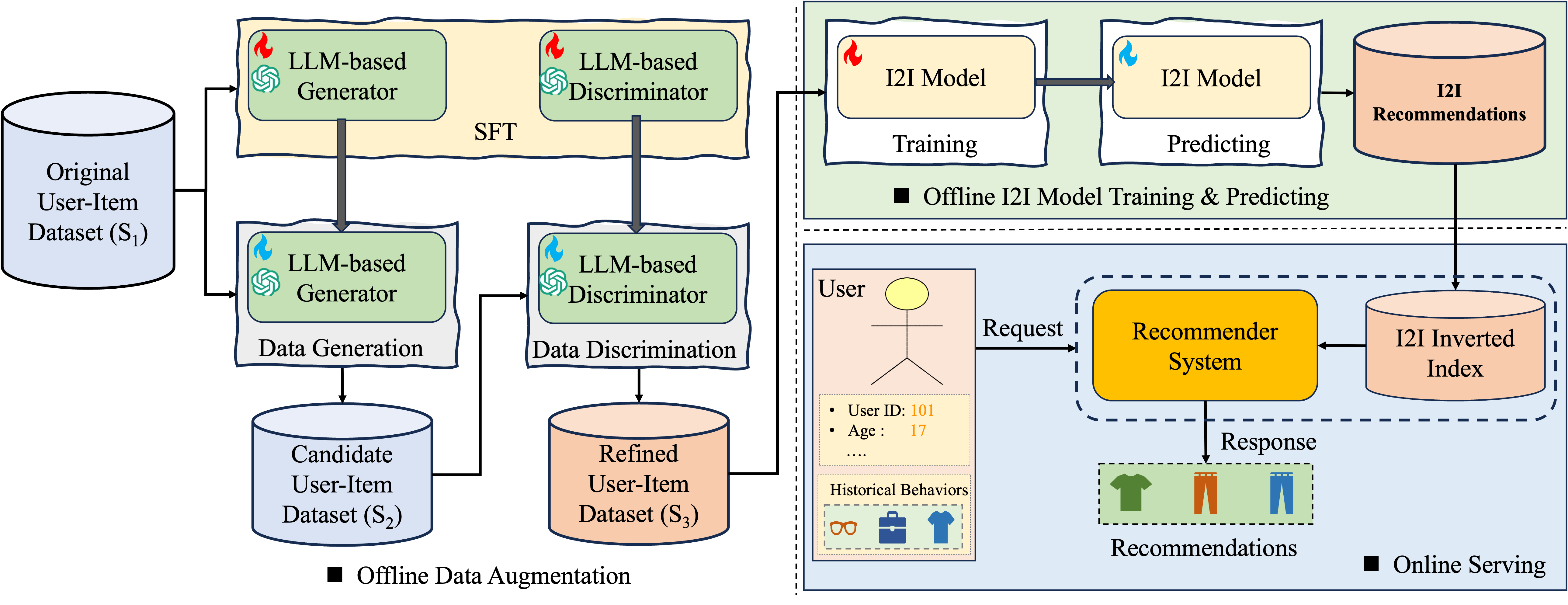}
    \caption{The framework of our proposed \textbf{LLM-I2I}.
    }
    \label{fig_overview}
\end{figure*}

\section{Related Work}
% \subsection{I2I Recommendation}
% Over the past decade, Item-to-Item (I2I) recommendation has seen significant advancements with various algorithms addressing its core challenges such as data sparsity, noise, cold start, and popularity bias\cite{linden2003amazon,zhang2025mixrec,wu2021self}. Here, we focus on data sparsity and noise issues closely related to this paper. Matrix factorization \cite{vozalis2005applying,ren2009collaborative} and self-supervised learning methods \cite{huang2022self,ren2023sslrec} have been developed to uncover latent factors or model data distribution to alleviate data sparsity problems. However, the high computational complexity prevents them from scaling to large datasets. The resampling and reweighting methods can modify the significance of various data points, thereby affecting the influence of noisy data on algorithm performance \cite{zhao2024data}. Optimizing model structure can further enhance the robustness of I2I models \cite{tian2022learning,jain2023sampling,he2024double}, but it requires customized modifications for different models, leading to high labour costs. Since many existing I2I models are still in use, optimizing these models using data-centric methods \cite{lai2024survey,bhatt2024data} presents a challenge with both commercial potential and application value.

% \subsection{LLM Enhanced Recommendation}
With the increasing research and application of LLMs, \textbf{L}arge \textbf{L}anguage \textbf{M}odel \textbf{E}nhanced \textbf{R}ecommender \textbf{S}ystems (LLM-ERS) has gained considerable attention recently \cite{liu2024large}. Traditional recommender systems can be enhanced using LLMs by aiding in training or generating synthetic data without LLM inference during online services. Existing LLM-ERS can be categorized into three main groups: \textit{Model Enhancement}, \textit{Knowledge Enhancement}, and \textit{Interaction Enhancement}. Model-enhanced LLM-ERS captures the semantic information and feeds it into the \textbf{R}ecommendation \textbf{S}ystem (RS) model\cite{li2023ctrl,wang2024flip}. Knowledge-enhanced LLM-ERS utilize LLMs to generate text descriptions about users and items. These descriptions are then used in recommendation system models as additional inputs for feature extraction \cite{liu2024once,luo2024large}. Interaction-enhanced LLM-ERS refines and generates interactions between users and items to tackle data sparsity and noise challenges\cite{huang2024large,wang2024llm4dsr}. Current data augmentation methods primarily focus on either data generation or data selection. However, they often neglect the integration of both strategies and fail to address potential noise issues in generated data. Unlike existing works, our proposed LLM-I2I integrates an LLM-based data generator with a data discriminator to effectively address data sparsity and noise issues. Furthermore, as demonstrated in the experimental section of this paper, denoising the generated data is crucial. 

\section{Proposed Methodology}
%\subsection{Framework Overview}
% We propose a \textbf{L}arge \textbf{L}anguage \textbf{M}odel enhanced \textbf{I}tem-to-\textbf{I}tem method (LLM-I2I) to overcome data sparsity and noise problems jointly. The framework of LLM-I2I is shown in Figure \ref{fig_overview}. Our framework consists of two key components: an LLM-based data generator and an LLM-based data discriminator. During the offline data augmentation, we first learn an LLM-based generator and discriminator using \textbf{S}upervised \textbf{F}ine-\textbf{T}uning (SFT) technique. Then, the LLM-based generator generates several user-item interactions containing the items the user may prefer. Subsequently, the LLM-based discriminator evaluates the original and synthesized data to obtain refined data. Next, using this data, we follow the existing I2I model training and prediction pipeline to train the I2I model and make predictions offline. Finally, with the I2I inverted index, the online recommender system can respond to users' online requests in real time. In what follows, we will provide more details about LLM-I2I.

% 去掉分节，节约版面。
%\subsection{Preliminary}
Given a user $u_i \in \mathcal{U}$ with historical behavior records (\eg the clicked or purchased item sequence) $\mathcal{Y}_{u_{i}}^{t} = \{Y_{u_{i}}^{1}, \cdots, Y_{u_{i}}^{t}\}$ at time $t$, the objective of recommender systems is to generate recommendations $\mathcal{Y}_{u_{i}}^{t+1}=\{Y_{u_{i}}^{t+1}, \cdots, Y_{u_{i}}^{t+m}\}$ to match the user's preference. To overcome the data sparsity problem, we need to learn an LLM-based generator parameterized by $\mathbf{\Theta}_{g}$ with user features that are denoted as $\mathbf{X}_{u_i}$ and a specific prompt instruction $\mathcal{P}_{g}$. Here, the user features consist of two parts: the static information (\ie the user ID and profile information) $\mathbf{X}_{u_i}^{s}$ and the dynamic information (\ie historical behavior records) $\mathbf{X}_{u_i}^{h^{t}}$. To refine the data, we learn an LLM-based discriminator parameterized by $\mathbf{\Theta}_{d}$ with the user features $\mathbf{X}_{u_i}$, and the item features $\mathbf{X}_{y_j}$, where $y_j \in \mathcal{I}_{u_{i}}$ denotes the evaluated item, and a specific prompt instruction $\mathcal{P}_{d}$.

\subsection{LLM-based Data Generator}
\begin{figure}
    \centering
    \includegraphics[width=1.0\linewidth]{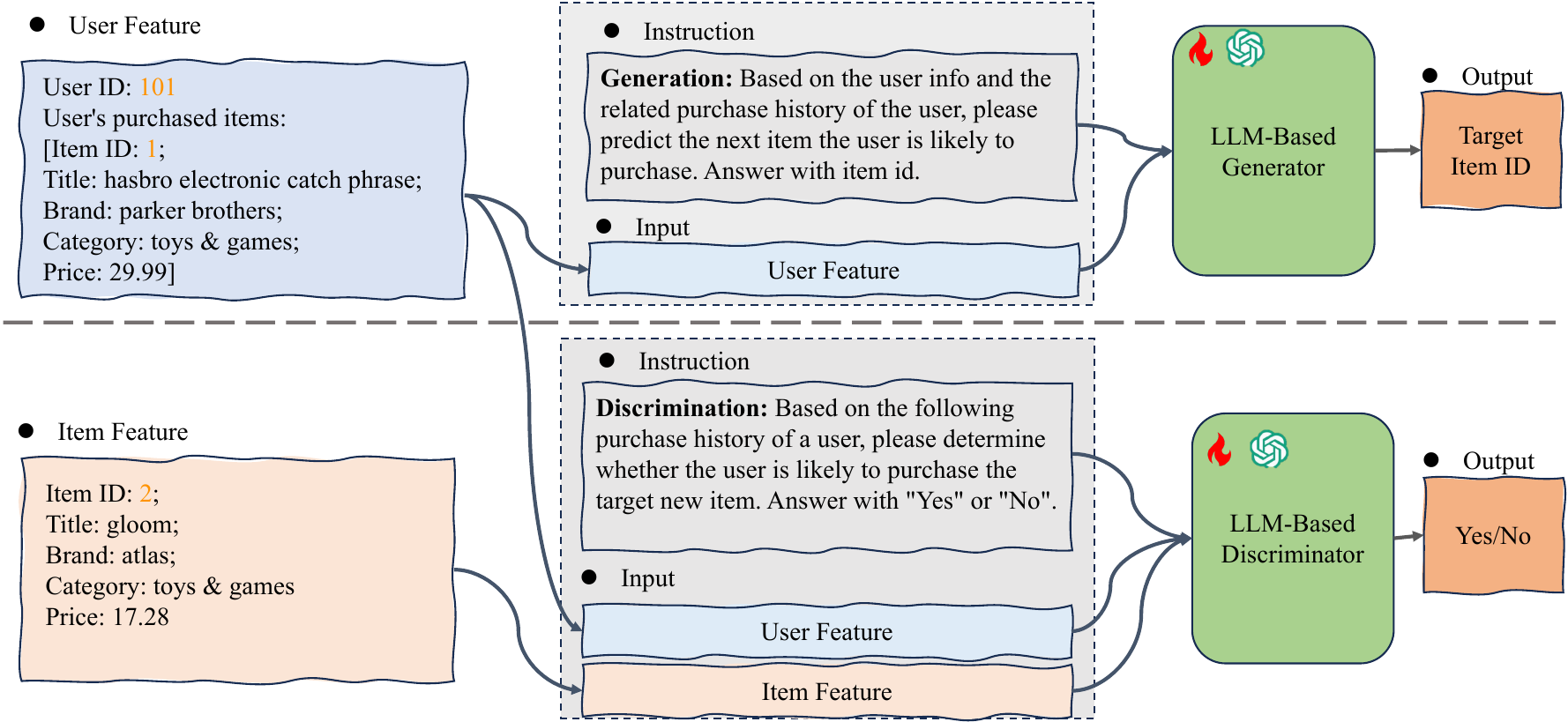}
    \caption{Supervised Fine-Tuning (SFT) of the LLM-based generator and discriminator.}
    \label{fig_generation_and_discrimination}
\end{figure}

\begin{figure}
    \centering
    \includegraphics[width=1.0 \linewidth]{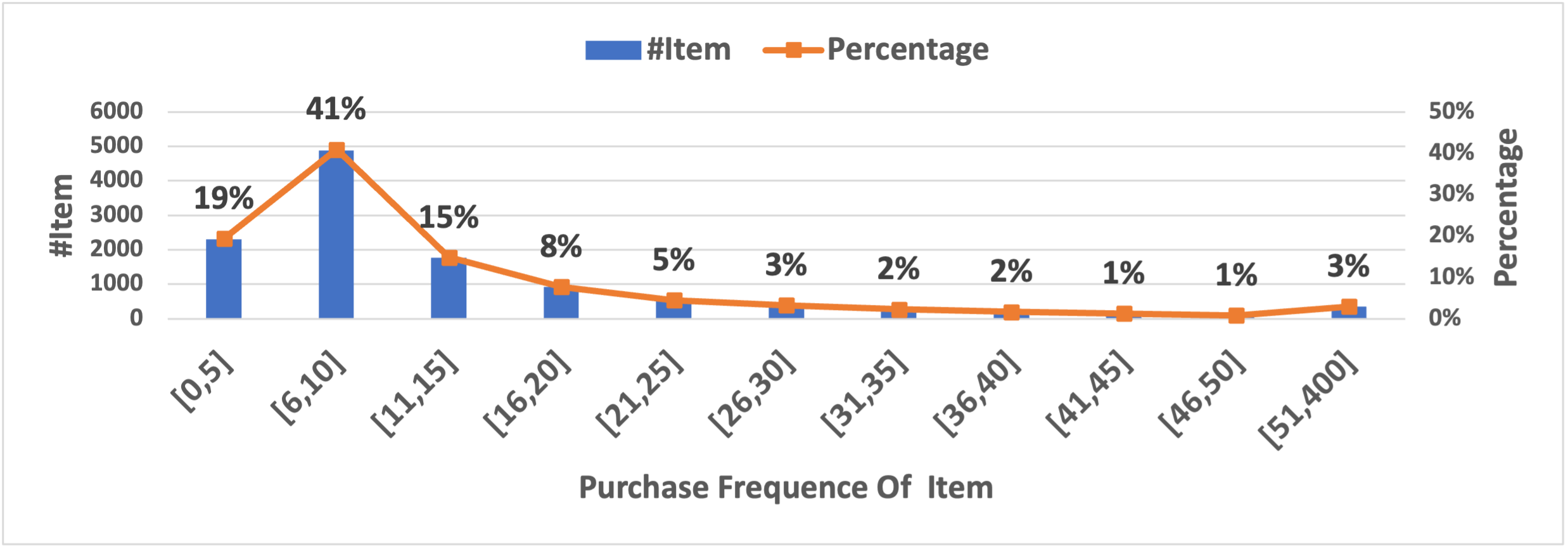}
    \caption{Data distribution in ARD Toys And Games dataset.}
    \label{fig_item_click_discribution}
\end{figure}

%% keypoint1: 大语言模型需要做微调，以适应电商场景；
Recently, LLMs have been known for their powerful natural language understanding and explanation generation abilities. However, directly using these pre-trained LLMs as data generators may not be optimal since these models are trained with general datasets, which may differ from a specific dataset, and their domain adaptability needs strengthening, especially in the e-commerce scenario. To this end, we adopt the \textbf{S}upervised \textbf{F}ine-\textbf{T}uning (SFT) technique to learn a LLM-based data generator. As shown in Figure ~\ref{fig_generation_and_discrimination}, given a well-designed generation prompt instruction $\mathcal{P}_{g}$ and the corresponding user features $\mathbf{X}_{u_i}$ as input, the LLM-based generator try to predict the target item ID that the user may prefer. Therefore, the SFT process can be formulated as below:
\begin{equation}
  \begin{aligned}
 	\mathcal{L}_{g} &= \sum_{u_{i}\in \mathcal{U}} \sum_{t=1}^{T} \textit{loss}(Y_{u_{i}}^{t+1} | \mathcal{P}_{g}, \mathbf{X}_{u_i}; \mathbf{\Theta}_{g}) \\
                &= \sum_{u_{i}\in \mathcal{U}} \sum_{t=1}^{T} \textit{loss}(Y_{u_{i}}^{t+1} | \mathcal{P}_{g}, (\mathbf{X}_{u_i}^{s}, \mathbf{X}_{u_i}^{h^{t}}); \mathbf{\Theta}_{g})\\
 \end{aligned}
 \label{eq:generation_loss}
\end{equation}
where $\textit{loss}(\cdot)$ is the loss function of LLMs, and $T$ equals the length of the user's historical behaviour records.

%% keypoint2: 在微调的过程中，需要对长尾商品进行重点关注，从而使得生成结构更多偏向长尾商品，从而补足长尾商品交互行为稀疏的问题。
% As we know, user behaviour data has a significant long-tail distribution characteristic in the e-commerce scenario. In Figure ~\ref{fig_item_click_discribution}, we can see that nearly 20\% of items have no more than five purchases in the Toys And Games of Amazon Review Dataset (ARD). 

%% keypoint3: 引入特殊loss，控制sft过程，对llm进行微调，从而得到我们想要的数据生成器。 
As we know, user behaviour data has a significant long-tail distribution characteristic in the e-commerce scenario. In Figure ~\ref{fig_item_click_discribution}, we can see that nearly 20\% of items have no more than five purchases in the Toys And Games of Amazon Review Dataset (ARD). In other words, these products rarely co-occur with other items. Thus, the classic I2I algorithms struggle to calculate the similarity between them and other items in such a data sparsity scenario and even fail to generate recommendations for these long-tail items. To overcome this problem, we propose a long-tail aware loss function to focus on learning the underlying patterns of long-tail items and improve the data generation ability of the LLM-based generator for long-tail items. We reformulate the objective function as below:
\begin{equation}
  \begin{aligned}
        \mathcal{L}_{g} &= \sum_{u_{i}\in \mathcal{U}} \sum_{t=1}^{T} \textbf{w}_{u_{i}}^{t+1} \textit{loss}(Y_{u_{i}}^{t+1} | \mathcal{P}_{g}, (\mathbf{X}_{u_i}^{s}, \mathbf{X}_{u_i}^{h^{t}}); \mathbf{\Theta}_{g}) \\
 	s.t. \quad \textbf{w}_{u_{i}}^{t+1} &=
 	\begin{cases}
 		\alpha & \textit{if \,$Y_{u_{i}}^{t+1}$ is the long-tail item,} \\
 		\beta & otherwise.
 	\end{cases} \\
 \end{aligned}
 \label{eq:long_tail_loss}
\end{equation}
where $\alpha$ and $\beta$ are two hyper-parameters for long-tail items and no-long-tail items, seperately. The greater value of $\alpha$ than $\beta$ could encourage the LLM-based generator to generate long-tail items as much as possible. We set $\alpha=4.0$ and $\beta=1.0$ in our experiments.

After training an LLM-based data generator, we feed it with the original user historical behavior data to synthesize a sequence of user-item interactions for each user. Then, we combine the original and synthesized user-item interactions into a candidate user-item dataset, as shown in Figure ~\ref{fig_overview}.

\subsection{LLM-based Data Discriminator}
%% key-point1: 在以往的研究中，数据合成后，直接或者经过简单的筛选，就用于后续的模型训练。这种处理方式会带来比较大的问题，一方面合成数据的质量存在不确定性，另外，在原始数据中添加不同比例的合成数据，会对最终模型训练数据分布产时影响，进而带来模型效果的差异。如我们实验测试显示，直接使用全部合成数据，

\begin{figure}[h]
		\centering
            \begin{subfigure}{0.48\linewidth} % 调整宽度为0.45\textwidth可以让两图并排
			\centering
			\includegraphics[width=\linewidth]{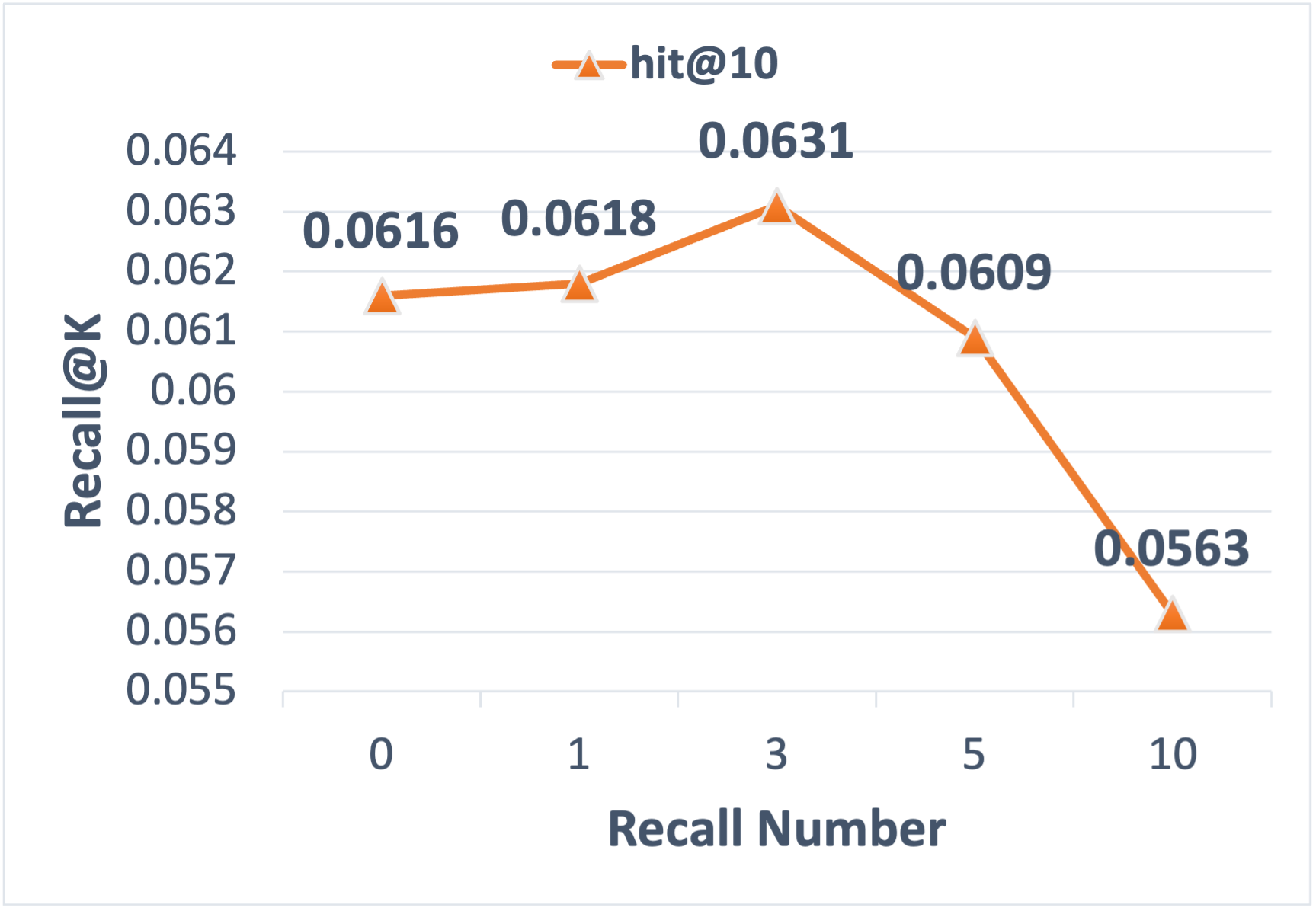} % 替换为你的图片路径
			%\caption{Recall@K for different recall numbers.}
			  \caption{The influence of recall number on the LLM-based generator.}
			\label{fig_recall_generation}
		\end{subfigure}\hfill
            \begin{subfigure}{0.48\linewidth}
			\centering
			\includegraphics[width=\linewidth]{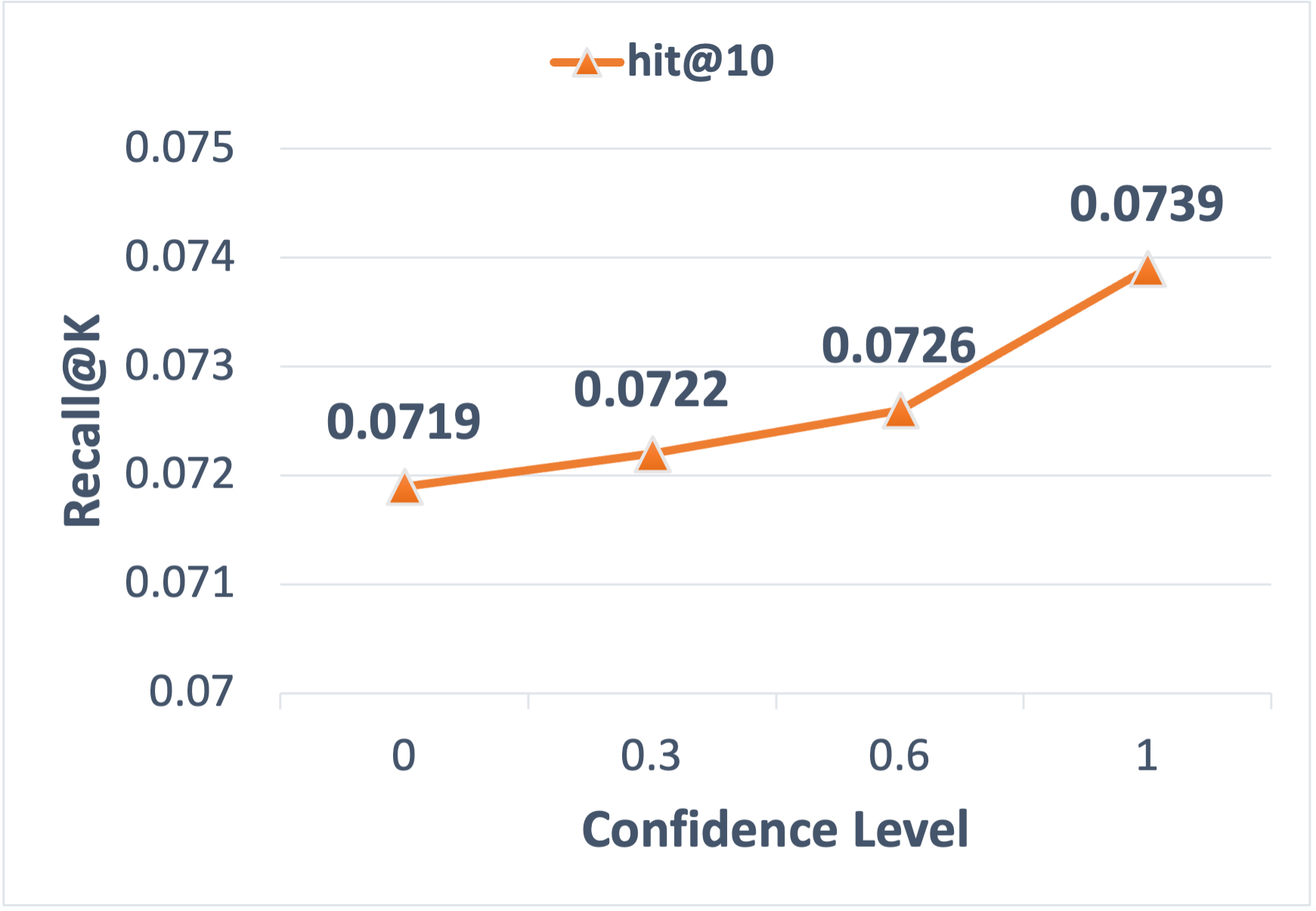} % 替换为你的图片路径
			%\caption{Recall@K for various confidence levels.}
			\caption{The effect of confidence level on the LLM-based discriminator.}
			\label{fig_recall_threshold}
		\end{subfigure}
		\caption{Recall@K under different settings for the LLM-based generator and discriminator.}
	\end{figure}

In previous studies, after data synthesis, it was directly or just filtered with certain rules made manually by experts and then used for subsequent algorithm model training. This simple processing approach is prone to two significant issues. Firstly, the quality of the synthetic data remains uncertain, and its direct incorporation into the training dataset may introduce noisy data, adversely affecting the model's training efficacy. Secondly, the distribution of the synthetic data may deviate from that of the original real data. Such a distribution drift can result in degrading the trained model's performance. To investigate these two issues, we evaluated the performance of a widely applied industrial I2I model named Swing~\cite{yang2020large} with different confidence levels (which were measured by a well-trained LLM-based discriminator) of training data and different recall numbers of synthesized user-item interactions for each user. We get two valuable experimental conclusions from Figure ~\ref{fig_recall_generation}. First, the higher the confidence level of the synthesized data, the better the training model will be. Thus, performing a data evaluation and selection on these synthesized data is very necessary. Second, as the amount of synthesized data increases, the performance of the training model will first increase and then decrease, which means that as the distribution of training data changes, the model performance may be worse.

To this end, we learn an LLM-based data discriminator to filter low-quality user-item interactions in synthesized data while maintaining high-quality training data, as illustrated in Figure~\ref{fig_overview}. The SFT processing for training the proposed discriminator can be formally expressed as below:
% \begin{equation}
% 	\begin{aligned}
%  	\mathcal{L}_{d} &= \sum_{u_{i}\in \mathcal{U}} \sum_{Y^{k} \in \mathcal{Y}_{u_{i}}^{+} \cup \mathcal{Y}_{u_{i}}^{-}} \textit{loss}(\textit{sgn}(u_{i}, Y^{k}) | \mathcal{P}_{d}, (\mathbf{X}_{u_i}^{s}, \mathbf{X_{u_i}^{h^{t}}), \mathbf{X}_{Y^{k}}; \mathbf{\Theta}_{d}) \\
%         & s.t. \quad \textit{sgn}(u_{i}, Y^{k}) =
%  	\begin{cases}
%  		1 & \textit{if \, $Y^{k} \in \mathcal{Y}_{u_{i}}^{+}$} \\
%  		0 & otherwise,
%  	\end{cases} 
%  \end{aligned}
% 	\label{eq:llm_discrimination_loss}
% \end{equation}

\begin{equation}
    \begin{aligned}
        \mathcal{L}_{d} &= \sum_{u_{i}\in \mathcal{U}} \sum_{Y^{k} \in \mathcal{Y}_{u_{i}}^{+} \cup \mathcal{Y}_{u_{i}}^{-}} \textit{loss}\left(\textit{sgn}(u_{i}, Y^{k}) \, \big| \, \mathcal{P}_{d}, \mathbf{X}_{u_i}, \mathbf{X}_{Y^{k}}; \mathbf{\Theta}_{d}\right) \\
        & \text{s.t.} \quad \textit{sgn}(u_{i}, Y^{k}) =
        \begin{cases}
            1 & \text{if } Y^{k} \in \mathcal{Y}_{u_{i}}^{+}, \\
            0 & \text{otherwise}.
        \end{cases}
    \end{aligned}
    \label{eq:llm_discrimination_loss}
\end{equation}
where $\mathcal{Y}_{u_{i}}^{+}$ and $\mathcal{Y}_{u_{i}}^{-}$ denote the sets of positive and negative interaction items for user $u_i$, respectively. Specifically, $\mathcal{Y}_{u_{i}}^{+}$ consists of the observed user-item interactions for $u_i$, while $\mathcal{Y}_{u_{i}}^{-}$ is constructed through random global negative sampling for $u_i$. Additionally, $\mathbf{X}_{Y^{k}}$ is the feature representation of item $Y^{k}$.

With a learned LLM-based discriminator, both the label and the corresponding probability, which reflects the confidence level, can be obtained. Figure ~\ref{fig_recall_threshold} illustrates the improvement in Swing when utilizing LLM-based generator outputs with varying confidence levels as synthetic data. The indicator Recall@10 is enhanced as the confidence level increases. This indicates that higher-quality augmented data leads to more accurate results. %Consequently, in the LLM-I2I framework, we employ the LLM-based discriminator for data filtering, retaining only those items classified as "Yes" with a confidence level of 1.0.

\subsection{Data Augmentation}
Once the LLM-based generator and discriminator have been fine-tuned, the former is employed to generate new data, while the latter is used to filter the data. The high-quality user-item pairs identified through this process are retained as the final synthetic data to enhance the I2I model. In recommendation systems, users' recent behaviours exert a stronger influence than their past behaviours \cite{sdm}. Additionally, selecting relevant sequences as input is critical due to the input length constraints of large language models. LLM-I2I utilizes each user's most recent ten user-item interactions as input during prediction to address this.

The items generated by the LLM-based generator are treated as target items. The LLM-based discriminator evaluates these items by considering the user's historical behaviours and preferences to predict whether the user will interact with them. The synthetic data used to train the I2I model comprises all predictions from the LLM-based generator that the discriminator classifies as "Yes" with a confidence score of 1.0. This enhanced data is then combined with the original data, which contains real user behaviours, to form the final refined training dataset for subsequent I2I model training, as illustrated in Figure~\ref{fig_overview}.

\subsection{Online Serving}
\label{module_online_serving}

% Based on the refined training dataset, we can follow the conventional I2I training and prediction pipeline to conduct offline model training and similar product recommendations, as shown in Fig.~\ref{fig_overview}. To obtain a real-time response online, we need to convert these recommendations into an I2I inverted index, which is a fast lookup table consisting of \textit{\<item_ID, recommendation_items\>} pairs. In order to ensure the retrieval speed and control the index size of the recommender system, the top $K$ recommendation items will be kept and stored in the engine. When providing recommendation services for a user, the recommendation system first queries the top $M$ item IDs associated with the user's most recent interactions. It then retrieves the recommendation results for these items from the inverted index. After aggregating and ranking these results, they are passed to the subsequent ranking pipeline for further sorting. Finally, the top-ranked items are selected and displayed to the user. 

Based on the refined training dataset, we adhere to the conventional I2I training and prediction pipeline to perform offline model training and generate similar product recommendations, as illustrated in Figure~\ref {fig_overview}. These recommendations are transformed into an I2I inverted index to enable real-time online responses—a high-performance lookup table composed of \textit{(Item\_ID, Recommendation\_Items)} pairs. In order to ensure rapid retrieval and manage the index size within the recommender system, only the top $K$ recommendation items are retained and stored in the engine.

When delivering recommendation services to a user, the system first queries the top $M$ item IDs corresponding to the user's most recent interactions. These IDs are then used to fetch recommendation results from the inverted index. Then, about $M\times K$ retrieved results are aggregated, ranked, and forwarded to the subsequent ranking pipeline for further refinement. Ultimately, the highest-ranked items are selected and presented to the user.

% Integrating LLM-I2I into an online recommender system only necessitates replacing its I2I inverted index table. Consequently, LLM-I2I offers significant practical advantages, as it preserves the existing I2I model architecture, avoids additional online deployment complexities, and minimizes resource demands. This makes it a highly cost-effective solution for many real-world applications.

\section{Experimental Setup}
\subsection{Datasets}
%% 为了充分证明LLM-I2I算法的有效性，我们在开源的学术研究数据集(Amazon datasets)和工业数据集(AEDS)分别进行了实验。他们的详细指标如表1所示。
%% Amazon Dataset 是最广泛使用的开源推荐基准之一，它记录了Amazon平台上不同类目下用户对于购买过的商品的评价。为了实验的公平性，我们与LLM-CF一样，都采用：运动与户外、美容、玩具与游戏 这三个数据集作为我们的评测集。同时，我们对于这三个数据集的训练、验证和测试集的划分方法与LLM-CF算法保持一致。此外，为了验证LLM-I2I能够提升长尾商品的算法效果，我们对每一个数据集涉及到的商品按照购买次数的多少从大到小排序，将排名靠后的20%的商品作为长尾商品。
%%https://jmcauley.ucsd.edu/data/amazon/links.html
%% AEDS 是一个来源于真实的大型电商搜索系统 AliExpress的十亿级的工业数据集。它包含了非常多的用户和当前用户历史点击过的商品。按照用户行为的时间顺序组织，我们将前90天的样本指定为训练集，并分别随机从后一天的样本中选择一百万的用户指定为验证集和测试集。经过分析，在AEDS数据集中，有将近25%的商品的点击次数仅为一次，我们将这部分商品作为该数据的长尾商品，从而验证LLM-I2I算法对于大规模实际的推荐系统中对于长尾商品的提升效果。

We evaluated our approach on a publicly available academic dataset (Amazon Review Dataset, ARD) and a proprietary industrial dataset (AliExpress Dataset, AEDS). Detailed statistics for both datasets are provided in Table~\ref{tab:dataset}.

\textbf{ARD}\footnote{https://jmcauley.ucsd.edu/data/amazon/links.html} is one of the most widely adopted public recommendation benchmarks, containing comprehensive user and item information, as well as user reviews for purchased items across diverse Amazon categories. Following the work~\cite{sun2024large}, we use the same three categories (\ie \textit{Beauty}, \textit{Sports and Outdoors}, and \textit{Toys and Games}) in our experiments. Additionally, to validate the effectiveness of LLM-I2I in addressing data sparsity, we sorted the items in each dataset by purchase frequency in descending order. Then we identified the bottom 20\% as long-tail items.

\textbf{AEDS} is a billion-scale industrial dataset collected from AliExpress.com\footnote{AliExpress.com is the official platform of AliExpress, a leading global B2C cross-border e-commerce retail platform under the Alibaba Group.}, a large-scale e-commerce B2C retail platform. The training set is constructed by chronologically organizing samples over a 90-day period. For the validation and test sets, one million users are randomly selected from the samples of the following day. A detailed analysis reveals that nearly 25\% of the items in AEDS were clicked only once, classifying them as long-tail items. This subset evaluates whether LLM-I2I can effectively mitigate data sparsity challenges in large-scale recommendation systems.

%% 数据集的大小
\begin{table}[htbp]
    \caption{Detailed statistics of the datasets. Here, LTI denotes long-tail items.}
    \begin{adjustbox}{max width=\linewidth}
    \label{tab:dataset}
    %            \begin{tabular}{cccccl}
    %     \toprule
    %         \multicolumn{2}{c}{Dataset} & $\#Users$ & $\#Items$ & $\#<U, I>$ & $\#LTI$\\
    %         \midrule
    %          \multirow{4}{*}{Beauty} & Training & 22,363 & 12,101 & 199,612 & 2,181  \\
    %          & Validation & 22,363 & 11,521 & 22,363 & 2,020 \\
    %           & Testing  & 22,363 & 11,457 & 22,363 & 1,986 \\
    %           & Total  & 22,363 & 12,101 & 244,338 & 2,181 \\
    %           \midrule
    %            \multirow{4}{*}{Sports And Outdoors} & Training & 35,598 & 18,357 & 296,099 & 3,286 \\
    %          & Validation & 35,598 & 17,567 & 35,598 & 3,039 \\
    %           & Testing  & 35,598 & 17,536 & 35,598 & 3,054 \\
    %           & Total  & 35,598 & 18,357 & 367,295 & 3,286 \\
    %            \midrule
    %            \multirow{4}{*}{Toys And Games} & Training & 19,412 & 11,924 & 171,165  & 2,310 \\
    %          & Validation & 19,412 & 11,184 & 19,412 & 2,077 \\
    %           & Testing  & 19,412 & 11,066 & 19,412 & 2,065 \\
    %           & Total  & 35,598 & 11,924 & 209,989 & 2,310 \\
    %           \midrule
    %           \multirow{4}{*}{AEDS} & Training & 209M & 68M & 10B & 16M\\
    %          & Validation & 1M & 4M & 5M & 1M\\
    %           & Testing  & 1M & 4M & 5M & 1M\\
    %           & Total  & 209M & 68M & 10B & 16M \\

    %     \bottomrule
           
    % \end{tabular}

    \begin{tabular}{ccccl}
        \toprule
            Dataset & $\#Users$ & $\#Items$ & $\#<U, I>$ & $\#LTI$\\
            \midrule
             Beauty & 22,363 & 12,101 & 244,338 & 2,181 \\
              \midrule
               Sports And Outdoors & 35,598 & 18,357 & 367,295 & 3,286 \\
               \midrule
              Toys And Games & 35,598 & 11,924 & 209,989 & 2,310 \\
              \midrule
              AEDS & 209M & 68M & 10B & 16M \\

        \bottomrule
           
    \end{tabular}
    
    \end{adjustbox}
\end{table}    

\begin{table*}[htbp]
	\centering
	\caption{Performance comparison across three categories of ARD. The highest performance values are highlighted in bold, while the second-best methods are indicated with an asterisk (*). The performance improvement ratio between LLM-I2I and the No Data Augmentation (No) method is denoted by $\Delta_p$.}
	\label{tab:amazon_main_result}
	\begin{adjustbox}{max width=\textwidth}
	\begin{tabular}{cc|cccc|cccc|cccl}
		\hline
		  \multirow{2}{*}{Backbone} & \multirow{2}{*}{Data Augmentation} & \multicolumn{4}{c|}{Beauty}    & \multicolumn{4}{c|}{Sports And Outdoors}      & \multicolumn{4}{c}{Toys And Games} \\\cline{3-14}          &       & Recall@5 & Recall@10 & NDCG@5 &NDCG@10 & Recall@5 & Recall@10 & NDCG@5 &NDCG@10 & Recall@5 & Recall@10 & NDCG@5 &NDCG@10 \\   
		   \hline

		    %  \multirow{3}[2]{*}{Cosine} & No    & 0.0060* & 0.0216* & 0.0025* & 0.0082* & 0.0028* & 0.0099* & 0.0012* & 0.0034* & 0.0081* & 0.0259* & 0.0033* & 0.0095* \\         
		    %  & LLM-CF & 0.0044 & 0.0152 & 0.0021 & 0.0055 & 0.0015 & 0.0058 & 0.0007 & 0.0021 & 0.0043 & 0.0149 & 0.0020 & 0.0054 \\          
		    %  & LLM-I2I & \textbf{0.0091} & \textbf{0.0237} & \textbf{0.0042} & 0.0082* & \textbf{0.0051} & \textbf{0.0106} & \textbf{0.0024} & \textbf{0.0041} & \textbf{0.0112} & \textbf{0.0272} & \textbf{0.0051} & \textbf{0.0098} \\    
		    % \hline  
		     
		    %  \multirow{3}[2]{*}{TFIDF} & No    & 0.0063* & 0.025*& 0.0026* & 0.0086* & 0.0031* & 0.0109* & 0.0013* & 0.0038* & 0.0087* & 0.0296*& 0.0035* & 0.0103* \\         
		    %  & LLM-CF & 0.0033 & 0.0153 & 0.0015 & 0.0053 & 0.0014 & 0.0062 & 0.0006 & 0.0022 & 0.0031 & 0.0146 & 0.0014 & 0.0051 \\          
		    %  & LLM-I2I & \textbf{0.0105} & \textbf{0.0269} & \textbf{0.0050} & \textbf{0.0101} & \textbf{0.0052} & \textbf{0.0123} & \textbf{0.0025} & \textbf{0.0048} & \textbf{0.0129} & \textbf{0.0310} & \textbf{0.0059} & \textbf{0.0117} \\   
		    %  \hline  
		      
		        \multirow{4}{*}{BM25} & No    & 0.0060*  & 0.0209*  & 0.0025  & 0.0073*   & 0.0031*  & 0.0094*  & 0.0013*  & 0.0035*  & 0.0092*  & 0.0258  & 0.0037* & 0.0096*  \\          
		       & LLM-CF & 0.0059 & 0.0178 & 0.0029* & 0.0067 & 0.0026 & 0.0080 & 0.0012 & 0.0029 & 0.0067 & 0.0182 & 0.0032 & 0.0069 \\          
		       & LLM-I2I & \textbf{0.0076} & \textbf{0.0210} & \textbf{0.0036} & \textbf{0.0078} & \textbf{0.0040} & \textbf{0.0099} & \textbf{0.0018} & 0.0035* & \textbf{0.0111} & \textbf{0.0271}* & \textbf{0.0051} & \textbf{0.0099} \\    
         & $\Delta_p$  & +26.67\% & +0.48\% & +44.00\% & +6.85\% & +29.03\% & +5.32\% & +38.46\% & +0.00\% & +20.65\% & +5.04\% & +37.84\% & +3.13\% \\
		      \hline

		        \multirow{4}{*}{BPR} & No    & 0.0062 & 0.0131 & 0.0034 & 0.0056 & 0.0049* & 0.0092* & 0.0026* & 0.0040* & 0.0074 & 0.0103 & 0.0037 & 0.0046 \\          
		        & LLM-CF & 0.0062 & 0.0134* & 0.0034 & 0.0057* & 0.0043 & 0.0079 & 0.0024 & 0.0035 & 0.0084* & 0.0147* & 0.0044* & 0.0064* \\          
		        & LLM-I2I & \textbf{0.0089} & \textbf{0.0184} & \textbf{0.0055} & \textbf{0.0085} & \textbf{0.0068} & \textbf{0.0132} & \textbf{0.0042} & \textbf{0.0062} & \textbf{0.0128} & \textbf{0.0210} & \textbf{0.0071} & \textbf{0.0098} \\  
          &  $\Delta_p$ & +43.55\% & +40.46\% & +61.76\% & +51.79\% & +38.78\% & +43.48\% & +61.54\% & +55.00\% & +72.97\% & +103.88\% & +91.89\% & +113.04\% \\
		       \hline  
		         
		         \multirow{4}{*}{YoutubeDNN} & No     & 0.0132 & 0.0241 & 0.0083* & 0.0118* & 0.0067 & 0.0124 & 0.0040 & 0.0058 & 0.0111 & 0.0194 & 0.0065 & 0.0092 \\          
		         %& LLM-CF & 0.016* & 0.0291* & 0.0099* & 0.0139* & \textbf{0.0112} & \textbf{0.0191} & \ 0.0069* & \textbf{0.0094} & 0.0165* & 0.0285* & 0.0099* & 0.0138* \\          
		         & LLM-CF & 0.0138* & 0.0257* & 0.0082 & 0.0120* 
		         & 0.0083* & 0.0147* & 0.0048* & 0.0064*  & 0.0137* & 0.0245* & 0.0084* & 0.0118* \\     
		         & LLM-I2I & \textbf{0.0161} & \textbf{0.0281} & \textbf{0.0096} & \textbf{0.0135} & \textbf{0.0092} & \textbf{0.0179} & \textbf{0.0057} & \textbf{0.0084} & \textbf{0.0200} & \textbf{0.0311} & \textbf{0.0124} & \textbf{0.0160} \\   
            &  $\Delta_p$ & +21.97\% & +16.60\% & +15.66\% & +14.41\% & +37.31\% & +44.35\% & +42.50\% & +44.83\% & +80.18\% & +60.31\% & +90.77\% & +73.91\% \\
		        \hline

		         \multirow{4}{*}{Swing} & No    & 0.0355* & 0.0506* & 0.0242* & 0.029* & 0.0236* & 0.0330* & 0.0169* & 0.0200*  & 0.0471* & 0.0616* & 0.0324* & 0.0372* \\          
		         & LLM-CF & 0.0326 & 0.0481 & 0.0212 & 0.0262 & 0.0211 & 0.0294 & 0.0135 & 0.0162 & 0.0453 & 0.0679 & 0.0291 & 0.0365 \\          
		         & LLM-I2I & \textbf{0.0379} & \textbf{0.0572} & \textbf{0.0259} & \textbf{0.0321} & \textbf{0.0280} & \textbf{0.0380} & \textbf{0.0196} & \textbf{0.0229} & \textbf{0.0522} & \textbf{0.0739} & \textbf{0.0343} & \textbf{0.0413} \\  
                 & $\Delta_p$ & +6.76\% & +13.04\% & +7.02\% & +10.69\% & +18.64\% & +15.15\% & +15.98\% & +14.50\% & +10.83\% & +19.97\% & +5.86\% & +11.02\% \\
	\hline
		
	\end{tabular}
\end{adjustbox}
\end{table*}

\subsection{Baseline}
The Item-to-Item (I2I) collaborative filtering algorithms can be categorized into five groups: neighborhood-based~\cite{koren2009matrix}, matrix factorization-based~\cite{koren2009matrix}, deep learning-based~\cite{xue2019deep}, graph learning-based~\cite{liu2022graph} and LLM-enhanced~\cite{sun2024large}. LLM-I2I is used as a data augmentation method for one classical algorithm from each category to demonstrate its ability to enhance the performance of diverse I2I algorithms.  

\textbf{Neighborhood-based} I2I identifies related items by computing similarity distances. The widely-used distance metric function BM25 is selected as the baseline method. 

\textbf{Matrix factorization-based} methods aim to decompose the user-item interaction/rating matrix into latent representations of users and items. We adopt BPR\cite{rendle2012bpr}, a model that enhances recommendation accuracy by maximizing the preference distance with items clicked and not clicked by the same user, as the baseline for matrix factorization-based I2I.

\textbf{Deep learning-based} I2I retrieval has gained significant attention with the advancement of deep learning techniques. We employ YoutubeDNN~\cite{covington2016deep}, a widely-used deep retrieval model, as our backbone. 
%YoutubeDNN learns an embedding for each item through a multi-layer neural network. In our experiments, the cosine similarity between item embeddings generated by YoutubeDNN is computed to identify related items for recommendation.

\textbf{Graph learning-based} approaches leverage graph connectivity for recommendations, representing users and items as nodes within a graph. Swing~\cite{yang2020large}, a widely adopted I2I retrieval method in industrial recommendation systems, computes item similarity using a bipartite graph.

\textbf{LLM-based} methods enhance traditional recommendation systems with large language models (LLMs). Among these, LLM-CF\cite{sun2024large} is a notable algorithm that improves recommendation performance by identifying similar samples within a constructed In-context Chain-of-Thought (COT) dataset.

\subsection{Implementation Details}
All experiments were conducted on the Tesla A100 platform using the Adam optimizer. The LLM-based generator and discriminator models were trained with a batch size 16 and a maximum input length of 1024. We employed full-parameter supervised fine-tuning (SFT) with a learning rate 5e-5 and a random seed of 42. We utilized DeepSpeed Zero2~\cite{zero-2} and FlashAttention2~\cite{dao2023flashattention} to accelerate training. For a fair comparison with LLM-CF\footnote{https://github.com/Jeryi-Sun/LLM-CF/tree/main \label{llm_cf}}, we selected LLama2-7B-Chat as the base LLM model. For BM25\footnote{https://github.com/benfred/implicit \label{implicit}}, the parameter $k$ was set to 20. In BPR\textsuperscript{\ref {implicit}}, the factor was set to 100. Following the setup of LLM-CF, the random seed for YoutubeDNN\textsuperscript{\ref {llm_cf}} was fixed in 2023, with both user and item embedding dimensions set to 64. For Swing\footnote{https://www.alibabacloud.com/help/zh/pai/use-cases/improved-swing-similarity-calculation-algorithm}, each user was limited to a maximum of 1,000 clicks.

\subsection{Evaluation Metrics}
We define Recall@K as the recall accuracy for the top K items, while NDCG@K measures the correctness of the ranking order among the retrieved K items. Additionally, six complementary metrics are employed to assess retrieval effectiveness in online A/B testing: RN, L2P, CTR, CVR, GMV, and Latency. RN represents the total number of items retrieved by the recall system. L2P denotes the percentage of users who purchased being exposed to the recommendations. CTR and CVR are abbreviations for click-through rate and conversion rate, respectively. GMV stands for gross merchandise volume, and Latency quantifies the processing time required for retrieval.

\begin{table*}[htbp]
	\centering
	\caption{Performance evaluation of LLM-I2I in AEDS. }
	\label{tab:ae_main_result}
	\begin{adjustbox}{max width=\textwidth}
		\begin{tabular}{cc|cccccc|cccccl}
			\hline
		Backbone & Data Augmentation & Recall@5 & Recall@10 & Recall@20 & Recall@50 & Recall@100 & Recall@200 & NDCG@5 &NDCG@10 &NDCG@20 & NDCG@50 & NDCG@100 & NDCG@200 \\
			\hline

			 % \multirow{2}[0]{*}{Cosine} & No & 0.0003  & 0.0736  & 0.0958  & 0.1223  & 0.1371  & 0.1483  & 0.0002  & 0.0412  & 0.0468  & 0.0521  & 0.0545  & 0.0561  \\      
			 % & LLM-I2I & \textbf{0.0005 } & \textbf{0.0802 } & \textbf{0.1052 } & \textbf{0.1472 } & \textbf{0.1734 } & \textbf{0.1908 } & \textbf{0.0003 } & \textbf{0.0435 } & \textbf{0.0497 } & \textbf{0.0581 } & \textbf{0.0624 } & \textbf{0.0648 } \\   
			 % \hline  
			 
			 %  \multirow{2}[0]{*}{TFIDF} & No & 0.0516  & 0.0758  & 0.0974  & 0.1271  & 0.1435  & 0.1557  & 0.0339  & 0.0417  & 0.0472  & 0.0531  & 0.0558  & 0.0575  \\     
			 %   & LLM-I2I  &   \textbf{0.0581 } & \textbf{0.0785 } & \textbf{0.1033 } & \textbf{0.1361 } & \textbf{0.1618 } & \textbf{0.1861 }  & \textbf{0.0352 } & \textbf{0.0433 } & \textbf{0.0493 } & \textbf{0.0586 } & \textbf{0.0630 } & \textbf{0.0656 } \\  
			 %   \hline  
			   
			     \multirow{3}{*}{BM25} & No & 0.0504  & 0.0565  & 0.0755  & 0.1034  & 0.1220  & 0.1356  & 0.0337  & 0.0318  & 0.0367  & 0.0423  & 0.0453  & 0.0472  \\      
			     & LLM-I2I & \textbf{0.0556 } & \textbf{0.0730 } & \textbf{0.0952 } & \textbf{0.1288 } & \textbf{0.1568 } & \textbf{0.1792 } & \textbf{0.0355 } & \textbf{0.0408 } & \textbf{0.0463 } & \textbf{0.0530 } & \textbf{0.0576 } & \textbf{0.0607 } \\   
         &  $\Delta_p$ & +10.32\% & +29.20\% & +26.09\% & +24.56\% & +28.52\% & +32.15\% & +5.34\% & +28.30\% & +26.16\% & +25.30\% & +27.15\% & +28.60\% \\
			     \hline  
			     
			      % \multirow{2}[0]{*}{ALS} & No & 0.0400  & 0.0005  & 0.0009  & 0.0034  & 0.0043  & 0.0054  & 0.0265  & 0.0003  & 0.0004  & 0.0009  & 0.0010  & 0.0012  \\     
			      %  & LLM-I2I & \textbf{0.0513 } & \textbf{0.0007 } & \textbf{0.0015 } & \textbf{0.0039 } & \textbf{0.0061 } & \textbf{0.0069 } & \textbf{0.0338 } & \textbf{0.0003 } & \textbf{0.0006 } & \textbf{0.0010 } & \textbf{0.0014 } & \textbf{0.0015 } \\   
			      %  \hline  

			      %   \multirow{2}[0]{*}{LMF} & No & 0.0037  & 0.0063  & 0.0122  & 0.0215  & 0.0342  & 0.0477  & 0.0022  & 0.0031  & 0.0046  & 0.0064  & 0.0085  & 0.0103  \\      
			      %   & LLM-I2I & \textbf{0.0044 } & \textbf{0.0071 } & \textbf{0.0118 } & \textbf{0.0234 } & \textbf{0.0364 } & \textbf{0.0541 } & \textbf{0.0026 } & \textbf{0.0035 } & \textbf{0.0046 } & \textbf{0.0069 } & \textbf{0.0090 } & \textbf{0.0115 } \\   
			      %   \hline  

			         \multirow{3}{*}{BPR} & No & 0.0034  & 0.0049  & 0.0076  & 0.0118  & 0.0165  & 0.0229  & 0.0022  & 0.0027  & 0.0034  & 0.0042  & 0.0049  & 0.0058  \\      
			         & LLM-I2I & \textbf{0.0052 } & \textbf{0.0095 } & \textbf{0.0150 } & \textbf{0.0271 } & \textbf{0.0403 } & \textbf{0.0557 } & \textbf{0.0032 } & \textbf{0.0046 } & \textbf{0.0060 } & \textbf{0.0083 } & \textbf{0.0105 } & \textbf{0.0126 } \\ 
            &  $\Delta_p$ & +52.94\% & +93.88\% & +97.37\% & +129.66\% & +144.24\% & +143.23\% & +45.45\% & +70.37\% & +76.47\% & +97.62\% & +114.29\% & +117.24\% \\
			         \hline

			            \multirow{3}{*}{YoutubeDNN} & No & 0.0447  & 0.0594  & 0.0817  & 0.1078  & 0.1264  & 0.1527  & 0.0311  & 0.0367  & 0.0400  & 0.0481  & 0.0510  & 0.0522  \\     
			            & LLM-I2I  & \textbf{0.0500 } & \textbf{0.0721 } & \textbf{0.0926 } & \textbf{0.1286 } & \textbf{0.1455 } & \textbf{0.1829 }
			            & \textbf{0.0332 } & \textbf{0.0410 } & \textbf{0.0483 } & \textbf{0.0542 } & \textbf{0.0618 } & \textbf{0.0639 } \\ 
                 &  $\Delta_p$ & +11.86\% & +21.38\% & +13.34\% & +19.29\% & +15.11\% & +19.78\% & +6.75\% & +11.72\% & +20.75\% & +12.68\% & +21.18\% & +22.41\% \\
			         \hline

			           \multirow{3}{*}{Swing} & No & 0.0490  & 0.0664  & 0.0907  & 0.1198  & 0.1396  & 0.1622  & 0.0361  & 0.0417  & 0.0478  & 0.0535  & 0.0568  & 0.0599  \\     
			            & LLM-I2I  & \textbf{0.0581 } & \textbf{0.0785 } & \textbf{0.1033 } & \textbf{0.1361 } & \textbf{0.1618 } & \textbf{0.1861 } & \textbf{0.0420 } & \textbf{0.0485 } & \textbf{0.0548 } & \textbf{0.0612 } & \textbf{0.0654 } & \textbf{0.0688 } \\
                              & $\Delta_p$  & +18.57\% & +18.22\% & +13.89\% & +13.61\% & +15.90\% & +14.73\% & +16.34\% & +16.31\% & +14.64\% & +14.39\% & +15.14\% & +14.86\% \\
			\hline  
		\end{tabular}
	\end{adjustbox}
\end{table*}

\begin{table*}[htbp]
	\centering
	\caption{Performance evaluation for long-tail items. }
	\label{tab:long_tail_result}
\begin{adjustbox}{max width= \linewidth}
		\begin{tabular}{cc|cccc|cccc|cccl}
			\hline

         \multirow{2}[1]{*}{Dataset}  &  \multirow{2}[1]{*}{Data Augmentation} &\multicolumn{4}{c|}{Total Item}  & \multicolumn{4}{c|}{Long-Tail Item}    & \multicolumn{4}{c}{No Long-Tail Item}   \\ 
         \cline{3-14}
        
        &   & Recall@5 & Recall@10 & NDCG@5&NDCG@10  & Recall@5 & Recall@10 & NDCG@5&NDCG@10 & Recall@5 & Recall@10 & NDCG@5 &NDCG@10 \\   
	\hline  
	
        \multicolumn{1}{c}{\multirow{3}[4]{*}{Beauty}} & No & 0.0355 & 0.0506 & 0.0242 & 0.029 & 0.0222 & 0.0276 & 0.0147 & 0.0165 & 0.0365 & 0.0522 & 0.0249 & 0.0299   \\      
        & LLM-I2I & \textbf{0.0379} & \textbf{0.0572} & \textbf{0.0259} & \textbf{0.0321} & \textbf{0.0209} & \textbf{0.0303} & \textbf{0.0151} & \textbf{0.0182} & \textbf{0.0391} & \textbf{0.0592} & \textbf{0.0266} & \textbf{0.0331} \\    
         & $\Delta_p$ & +6.76\% & +13.04\% & +7.02\% & +10.69\% & -5.86\% & +9.78\% & +2.72\% & +10.30\% & +7.12\% & +13.41\% & +6.83\% & +10.70\%   \\
	\hline

        \multirow{3}[4]{*}{Sports And Outdoors}& No & 0.0236 & 0.033 & 0.0169 & 0.02 & 0.0107 & 0.0143 & 0.0069 & 0.0081 & 0.0246 & 0.0345 & 0.0177 & 0.0209   \\     
         &LLM-I2I & \textbf{0.028} & \textbf{0.038} & \textbf{0.0196} & \textbf{0.0229} & \textbf{0.0119} & \textbf{0.0175} & \textbf{0.0087} & \textbf{0.0105} & \textbf{0.0292} & \textbf{0.0395} & \textbf{0.0205} & \textbf{0.0238}   \\
          &  $\Delta_p$  & +18.64\% & +15.15\% & +15.98\% & +14.50\% & +11.21\% & +22.38\% & +26.09\% & +29.63\% & +18.70\% & +14.49\% & +15.82\% & +13.88\% \\
         \hline 
  
         \multirow{3}[3]{*}{Toys And Games} & No & 0.0471 & 0.0616 & 0.0324 & 0.0372 & 0.0273 & 0.0378 & 0.0192 & 0.0226 & 0.0489 & 0.0638 & 0.0336 & 0.0385  \\    
           & LLM-I2I & \textbf{0.0522} & \textbf{0.0739} & \textbf{0.0343} & \textbf{0.0413} & \textbf{0.0329} & \textbf{0.0434} & \textbf{0.0214} & \textbf{0.0249} & \textbf{0.0539} & \textbf{0.0767} & \textbf{0.0355} & \textbf{0.0428}   \\
               &   $\Delta_p$ & +10.83\% & +19.97\% & +5.86\% & +11.02\% & +20.51\% & +14.81\% & +11.46\% & +10.18\% & +10.22\% & +20.22\% & +5.65\% & +11.17\%   \\   
       \hline

       \multirow{3}[0]{*}{AEDS} & No& 0.0490 & 0.0664 & 0.0361  & 0.0417  & 0.0134 & 0.0186 & 0.0081 & 0.0098 & 0.0528 & 0.0715 & 0.0391 & 0.0451   \\      
       & LLM-I2I & \textbf{0.0581} & \textbf{0.0785} & \textbf{0.0420} & \textbf{0.0485} & \textbf{0.0227} & \textbf{0.0299} & \textbf{0.016} & \textbf{0.0182} & \textbf{0.0619} & \textbf{0.0837} & \textbf{0.0448} & \textbf{0.0518}  \\      
       & $\Delta_p$   & +18.57\% & +18.22\% & +16.34\% & +16.31\% & +69.40\% & +60.75\% & +97.53\% & +85.71\% & +17.23\% & +17.06\% & +14.58\% & +14.86\%  \\ 
        \hline 
		\end{tabular}
	\end{adjustbox}
\end{table*}

\section{Experimental Results}
%% 我们将通过回答以下几个问题来进行我们的实验：
% 1. 采用了LLM-I2I数据增强算法后，不同的I2I算法在算法指标上是否都能够得到增强？
% 2. 相比于现有的基于LLM的协同过滤增强算法，LLM-I2I算法的效果是否更好？
% 2. 采用了LLM-I2I数据增强算法后，长尾商品的算法指标是否能够得到提升？
% 3. LLM-I2I算法中哪些部分对于算法的影响最大？
% 4. 选择不同参数规模和模型结构的大语言模型是否会对算法结果有影响？
% 5. LLM-I2I部署到实际的在线系统后对业务指标会带来哪些提升？ 

% This section presents extensive experiments on ARD and AEDS datasets to address the following research questions:
% \begin{itemize}
% \item \textbf{RQ1:} Does LLM-I2I enhance the performance of different I2I algorithms?
% \item \textbf{RQ2:} How does LLM-I2I compare with state-of-the-art LLM-based collaborative filtering methods (\eg, LLM-CF)?
% \item \textbf{RQ3:} Can LLM-I2I improve metrics for long-tail items?
% \item \textbf{RQ4:} Which LLM-I2I component contributes most to its overall performance?
% \item \textbf{RQ5:} How does the choice of LLM architectures and parameter sizes impact the results?
% \item \textbf{RQ6:} What performance improvements can be expected when deploying LLM-I2I in real-world RS?
% \end{itemize}

\subsection{Performance Evaluation}
Table~\ref{tab:amazon_main_result} presents the performance comparison before and after applying LLM-I2I to various I2I algorithms on the ARD dataset. The results demonstrate significant improvements in Recall@5 and Recall@10 across all evaluated algorithms. Among the baseline methods, Swing achieves the highest Recall@10 scores in all ARD categories. LLM-I2I further enhances Swing's performance, yielding relative improvements of 13.04\%, 15.15\%, and 19.97\% in Recall@10 for the Beauty, Sports, and Toys categories, respectively.

To validate LLM-I2I's effectiveness, we conduct comparative experiments against LLM-CF, a representative LLM-based enhancement approach. The experimental results consistently show that LLM-I2I outperforms LLM-CF across different I2I algorithms. This performance advantage can be attributed to the fundamental architectural difference between the two approaches: while LLM-CF employs user-based enhancement, LLM-I2I adopts an item-based strategy. Our findings align with established literature \cite{deshpande2004item}, which demonstrates the general superiority of item-based collaborative filtering over user-based approaches.

To further evaluate the practical effectiveness of LLM-I2I in real-world recommendation systems, we conducted extensive experiments on AEDS, a large-scale industrial dataset. As demonstrated in Table~\ref{tab:ae_main_result}, LLM-I2I consistently enhances the performance of various I2I algorithms across both Recall@K and NDCG@K metrics. Notably, when applied to Swing, the best-performing offline algorithm, LLM-I2I achieves a remarkable $18.57\%$ and $16.34\%$ improvements in Recall@5 and NDCG@5, respectively. Based on these compelling offline results, we deployed the LLM-I2I-enhanced Swing algorithm as the core I2I indexing component in our production recommendation system.

\subsection{Long-tail Item Enhancement}
The long-tail effect in recommendation systems manifests when popular items with abundant user interactions dominate recommendations, often at the expense of higher-quality but less popular items. This phenomenon is particularly pronounced in I2I algorithms that rely on implicit feedback. To investigate LLM-I2I's capability in addressing this challenge, we employ Swing as our baseline algorithm.

As presented in Table \ref{tab:long_tail_result}, LLM-I2I demonstrates consistent improvements in both Recall and NDCG metrics for long-tail items across both academic and industrial datasets. Notably, the performance gains are more substantial on large-scale industrial datasets compared to smaller academic benchmarks. Specifically, on the AEDS dataset, LLM-I2I achieves remarkable improvements of 60.75\% and 85.71\% in Recall@10 and NDCG@10 for long-tail items, respectively. These results suggest that LLM-I2I's effectiveness in handling long-tail recommendations scales favorably with dataset size and complexity.

\subsection{Ablation Study}
To systematically evaluate the contribution of each component in LLM-I2I, we conducted an ablation study on the Toys and Games category of the ARD dataset. We designed six variants of LLM-I2I: 1)\textit{w/o all}: Baseline without any LLM-I2I components. 2)\textit{w/o LLM-based generator}: Utilizes only the LLM-based discriminator for raw data filtering. 3)\textit{w/o long-tail loss}: Removes the long-tail optimization objective from the generator. 4)\textit{w/o LLM-based discriminator}: Accepts all generator outputs as synthetic data. 5)\textit{w/o threshold filter}: Considers all discriminator "Yes" classifications as valid. 6)\textit{w/o Swing}: Uses only the generator's predictions without collaborative filtering.

% \begin{itemize}
%     \item \textit{w/o all}: Baseline without any LLM-I2I components.
%     \item \textit{w/o LLM-based generator}: Utilizes only the LLM-based discriminator for raw data filtering.
%     \item \textit{w/o long-tail loss}: Removes the long-tail optimization objective from the generator.
%     \item \textit{w/o LLM-based discriminator}: Accepts all generator outputs as synthetic data.
%     \item \textit{w/o threshold filter}: Considers all discriminator "Yes" classifications as valid.
%     \item \textit{w/o Swing}: Uses only the generator's predictions without collaborative filtering.
% \end{itemize}

%As evidenced by the experimental results in Table \ref{tab:ablation_study}, each component contributes significantly to the overall performance. Our analysis reveals several key insights: the LLM-based generator significantly improves recommendation quality through effective synthetic data generation, while the long-tail loss function specifically enhances performance on rare items. The LLM discriminator plays a critical role in maintaining data quality by filtering out noisy synthetic data. Furthermore, the integration of generative modeling with collaborative filtering demonstrates superior performance compared to standalone approaches, highlighting the value of combining these methodologies.

The experimental results in Table \ref{tab:ablation_study} show that each component contributes significantly to the overall performance. These findings lead to several key insights:
% \begin{enumerate}
\begin{itemize}
    \item The LLM-based generator's synthetic data generation capability substantially enhances recommendation quality.
    \item The long-tail loss function effectively boosts performance for underrepresented items.
    \item The LLM-based discriminator plays a crucial role in filtering noisy synthetic data.
    \item The integration of generative modelling with collaborative filtering (Swing) yields superior results compared to standalone approaches.
\end{itemize}
% \end{enumerate}

\begin{table}[htbp]
	\caption{Results of ablation study.}
	\label{tab:ablation_study}
		\begin{adjustbox}{max width=\linewidth}
	\begin{tabular}{lcccl}
		 \toprule
	Method & Recall@5 & Recall@10 & NDCG@5 & NDCG@10 \\
		\midrule
	LLM-I2I & \textbf{0.0522} & \textbf{0.0739} & \textbf{0.0343} & \textbf{0.0413} \\   
	
	w/o all & 0.0471 & 0.0616 & 0.0324 & 0.0372 \\
	 w/o LLM-based generator & 0.0468 & 0.0616  & 0.0319 & 0.0367  \\

   w/o long-tail loss & 0.0455 & 0.0653  & 0.0305 & 0.0369 \\   
   
    w/o LLM-based discriminator & 0.0454 & 0.0631 & 0.0305 & 0.0362  \\    
    
    w/o threshold filter & 0.0506 & 0.0719 & 0.0332 & 0.0400  \\ 
    w/o Swing & 0.0674 & 0.0721  & 0.0584 & 0.0610  \\ 
		
		\bottomrule
		
	\end{tabular}
\end{adjustbox}
	
\end{table}

\subsection{Online Evaluation Results}
To validate LLM-I2I's effectiveness in real-world scenarios, we implemented it in AliExpress' production search system. Following the architecture detailed in Section \ref{module_online_serving}, we maintained an I2I inverted index containing 200 related items per product. For query processing, we utilized users' 100 most recent clicks as search keys to ensure sufficient product retrieval. We conducted a rigorous one-week A/B test with the following experimental design: 4.0\% of users who received recommendations from the baseline Swing algorithm trained solely on historical user behavior. In contrast, an additional 4.0\% of users received recommendations from Swing enhanced with LLM-I2I. As presented in Table \ref{tab:online_metrics}, LLM-I2I demonstrates significant improvements across all key business metrics. These results substantiate LLM-I2I's practical value in large-scale e-commerce search systems. Notably, the implementation achieves these improvements without increasing system latency, as it operates through offline updates to the I2I inverted index. The demonstrated performance gains, coupled with the system's computational efficiency, highlight LLM-I2I's potential for real-world deployment in production recommendation systems.
\begin{table}[htbp]
	\caption{Online A/B experimental results.}
	\label{tab:online_metrics}
		\begin{adjustbox}{max width=\linewidth}
	\begin{tabular}{ccccccl}
		\toprule
		  RN & L2P & CTR & CVR  & GMV & Latency  \\
		\midrule
	+6.02$\%$ & +0.80$\%$ & +0.25$\%$& +0.11$\%$ & +1.22$\%$ & +0.37$\%$ \\
		\bottomrule
	\end{tabular}
\end{adjustbox}
\end{table}

\section{Conclusion}
We propose LLM-I2I to address the data sparsity and noise problems in traditional I2I models. Our approach uniquely integrates an LLM-based generator to predict user-item interactions and a discriminator to filter high-quality synthetic data. The generator incorporates a specific long-tail loss to better capture patterns in long-tail items. Experiments demonstrate LLM-I2I's effectiveness across industrial and academic datasets. Online A/B testing on AliExpress confirms significant improvements in key business metrics (RN and GMV).

%\bibliography{aaai2026}
\bibliography{main_v1}
% \balance
% \normalem   %% 解决参考文献里面期刊下划线问题


% \end{sloppypar}
\end{document}